\documentclass[12pt]{iopart}
\usepackage{iopams}
\usepackage{graphicx}

\begin{document}
\title{Rotational-state purity of Stark-decelerated molecular beams}
\author{N J Fitch,  D A Esteves, M I Fabrikant, T C Briles, Y Shyur, L P Parazzoli, and H J Lewandowski}
\address{JILA and Department of Physics, University of Colorado, 440 UCB, Boulder, CO 80309-0440}
\date{\today}

\begin{abstract}
Cold, velocity-controlled molecular beams consisting of a single quantum state promise to be a powerful tool for exploring molecular scattering interactions.  In recent years, Stark deceleration has emerged as one of the main methods for producing velocity-controlled molecular beams. However, Stark deceleration is shown not to be effective at producing a molecular beam consisting of a single quantum state in many circumstances.  Therefore, quantum state purity must be carefully considered when using Stark decelerated beams, particularly in collision experiments where contributions from all quantum states must be addressed.
\end{abstract}

\maketitle

\section{Introduction}

Molecular beams have been the workhorse for studying gas-phase interactions since the 1950's \cite{Taylor1955,Casavecchia2000}.  A recent goal of these studies has been to measure and possibly even control interactions on a state-by-state level, allowing exploration of the role of quantum mechanics in molecular processes. A prerequisite for these studies is the ability to prepare the reactants with well defined internal (electronic, rovibrational) and external (translational) degrees of freedom.  Examples of experiments that exploit restricted degrees of freedom include controlled chemistry \cite{Zare1998,Crim1996,Gordon1997}, stereochemistry \cite{Orr1996,deMiranda2011}, fundamental studies of bimolecular reaction dynamics \cite{Lee1987,Polanyi1987}, and scattering resonances \cite{Zare2006,Qiu2006}.  In particular, studies of scattering resonances benefit enormously from the use of beams with narrow velocity distributions \cite{Parazzoli2009,Bethlem2000} where the collision energy can be varied over the resonance \cite{Gilijamse2006}.  The study of \emph{cold} molecules \cite{Ye2009} is predicted to be particularly insightful for exploring intermolecular interactions and chemistry \cite{Softley2009,Krems2008}.  For example, increased control of the participating quantum states can elucidate the role of barriers in molecular reactions where ground-state interactions may be energetically forbidden but are possible among excited states.

Several methods have been developed to produce cold molecular samples with limited internal and external degrees of freedom.  These methods include supersonic expansion \cite{Scoles1988}, buffer-gas cooling \cite{Weinstein1998,Patterson2007}, velocity selection \cite{Rangwala2003}, kinematic cooling \cite{Elioff2003}, photoassociation \cite{Sage2005,Wang2004}, Feschbach resonances \cite{Ni2008,Danzl2010}, and beam deceleration using optical \cite{Fulton2006}, magnetic \cite{Narevicius2008} or electric fields \cite{Bethlem1999}. One such beam deceleration technology, Stark deceleration, uses inhomogeneous time-varying electric fields to decelerate a beam of polar molecules produced via supersonic expansion.  This method has been used to decelerate a variety of molecular species, including CO, NH$_3$, ND$_3$, OH, SO$_{2}$, NH, and CaF \cite{Bethlem1999,Bethlem2002,Bochinski2003,Meijer2006,Bucicov2008, Wall2011} and has been employed in a number of cold collision experiments \cite{Gilijamse2006,Parazzoli2011,Sawyer2011,Sawyer2008,Kirste2010,Scharfenberg2010,Scharfenberg2011}.

There are numerous experiments that require samples to be both translationally cold and in a well-defined single internal quantum state. Having proven to be highly effective at producing cold molecules, Stark deceleration is a logical candidate for the production of these cold molecular beams consisting of a \emph{single} state. Therefore, the physical chemistry community would benefit from a systematic study specifically addressing the state selectivity of Stark deceleration.

Explorations of Stark deceleration selectivity have appeared previously in the literature. One study focused on the deceleration of different isotopologues \cite{Bethlem2002} while another involved different projections of the dipole moment of the OH molecule onto the electric field axis \cite{vandeMeerakker_thesis}. Both of these studies explored selectivity among species with different Stark shift to mass ratios, which is the parameter that determines the dynamics of the deceleration process.  The present investigation expands on this previous work with a rigorous and detailed experimental study of the state selectivity of Stark deceleration.  The principal method used for this investigation is the deceleration and detection of various rotational states of a single isotopologue of ammonia. Specifically, molecules in two rotational states, which have different Stark shifts, are decelerated and state-selectively detected using decelerator timing sequences corresponding to a large range of Stark shift to mass ratios. These measurements are compared to the results of a series of three-dimensional Monte Carlo simulations to confirm the validity of the computational model. This model is then used to extract the final mean velocity for different rotational states decelerated under various conditions. We find that for a given timing sequence, the final mean velocity varies with rotational state.  This velocity is an important parameter for many collision experiments where collision energies must be precisely characterized.

\section{Stark Deceleration}
The Stark deceleration process begins with a pulsed supersonic expansion, where the molecules of interest are seeded as a low concentration in a heavier buffer gas.  The resulting molecular beam is internally cold with molecules predominantly in the rovibrational ground state.  However, if the molecule has multiple low lying, energetically accessible excited rotational states, as is the case with ammonia, a significant portion of the population can end up in these excited states.  For molecules such as OH, where the first excited rotational state is $\sim 80$ cm$^{-1}$ above the ground state,  only the ground state will be substantially populated and rotational excitations can usually be neglected.  Though internally cold, such beams have a high forward velocity in the laboratory frame of many hundreds of meters per second, which can be reduced by Stark deceleration.

The Stark deceleration technique \cite{Bethlem1999} relies on the interaction of an electric field with a molecule's electric dipole moment to reduce the high forward velocity a portion of the molecular beam.  A series of high-voltage electrode pairs create periodic electric field maxima along the beam axis.  Any molecule whose energy increases with increasing electric field (known as a weak-field-seeking state) will experience an acceleration opposing its motion.  As these molecules propagate into the increasing electric field, longitudinal kinetic energy is converted into potential energy.  If the molecules were allowed to continue past the local electric field maximum they would regain their lost kinetic energy.  However, before they reach the field maximum, the electric field is switched off rapidly, reducing the forward velocity of the beam by a small amount.  This process is repeated using successive electrode pairs to produce a molecular beam with the desired final velocity.  Transverse guiding is also achieved because the molecules are attracted to the electric field minimum along the center of the decelerator.  To guide equally in both transverse dimensions, successive electrode pairs are oriented orthogonally to one another.

The final velocity of the decelerated molecules is determined by the amount of energy removed per stage and the total number of stages.  The amount of energy removed per stage can be described by the position of the so-called ``synchronous molecule", which is at the same physical location relative to each electrode pair at the time the electric fields are switched off.  This position is referred to as the phase angle, $\phi_{0}$, where $\phi_{0}=0^{\circ}$ corresponds to no energy being removed and $\phi_{0}=90^{\circ}$ corresponds to the maximum amount of energy being removed.  Increasing the phase angle, and hence the amount of energy removed per stage, results in both lower final velocities and lower overall molecule numbers \cite{Osterwalder2010}.  This is because the phase-space acceptance, the region of phase space that will be decelerated and transported to the end of the decelerator, decreases with increasing phase angles.

\begin{figure}
\includegraphics[width = 8.6cm]{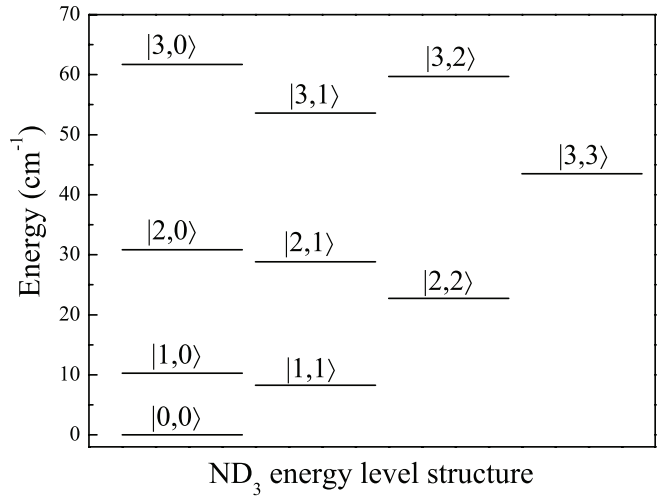}
\caption{Rotational energy levels of the ground electronic state in $^{14}$ND$_3$. The states are labeled by $|$\textit{J},\textit{K}$\rangle$ and organized by \textit{K} value. The inversion splitting is not visible on this scale.}
\label{energy}
\end{figure}

Molecules close to the synchronous molecule in phase space experience an average force toward its location.  This force results in an effective time-averaged moving potential well \cite{Bethlem2002}.  The dynamics of molecular trajectories in this moving potential well have been thoroughly investigated in the case where the molecules have the same mass and Stark shift as that of the synchronous molecule \cite{Meijer2005}. In the present analysis, molecules with the same mass but different Stark shifts are experimentally investigated.  This required a molecular beam with many initially populated weak-field-seeking states.  The numerous low-lying rotational states of the ammonia molecule (Figure \ref{energy}) are ideal for this purpose because of their appreciable Boltzmann factors at our beam temperature ($6-7$ K).

\section{The Ammonia molecule}

In its rovibrational ground state, ammonia has a trigonal pyramidal structure with a ladder of rotational states denoted by $|$\textit{J},\textit{K}$\rangle$, where \textit{J} is the total angular momentum and \textit{K} is the projection of \textit{J} onto the symmetry axis of the molecule. The trigonal pyramidal structure has two inversion eigenstates, which are symmetric and antisymmetric combinations of the two polarized ``umbrella" states. The umbrella states are defined by the nitrogen being either ``above" or ``below" the plane defined by the hydrogen atoms.  Because there is a barrier for the nitrogen to tunnel through this plane, the two eigenstates are separated in energy by an inversion splitting, $\Delta U_{inv}$, which varies only slightly among low rotational states (Table \ref{stark}).  When an electric field is applied, the two opposite-parity inversion states mix and are repelled from one another in energy.  This Stark energy is given by

\begin{equation}
\textit{U}_{stark} = \pm\sqrt{\left(\frac{\Delta U_{inv}}{2}\right)^2+\left(\left|\vec{\mu}\right|\left|\vec{\textit{E}}\right|\frac{M K}{J(J+1)}\right)^2} \mp \frac{\Delta U_{inv}}{2},
\end{equation}

where $\vec{\mu}$ is the electric dipole moment ($\mu_{ND_{3}}=1.47$ Debye), $\vec{E}$ is the electric field, $M$ is the projection of $J$ onto the electric-field axis, and the upper (lower) signs correspond to weak- (strong-) field-seeking molecular states. Fully deuterated ammonia (ND$_{3}$) was chosen over NH$_{3}$ because of its favorable Stark shift due to its much smaller inversion splitting ($\sim$ 0.05 cm$^{-1}$ compared to $\sim$ 0.79 cm$^{-1}$ for NH$_3$).

\begin{table}
\caption{The energies, inversion splittings, and relative dipole moments for various rotational levels.  Entries in the fourth column include all possible non-zero values of $M$.\label{stark}}
\begin{indented}
\item[]
\begin{tabular}{c c c c}
\hline
$|J,K\rangle$& E$_{rot}$ (cm$^{-1}$) &$\Delta$U$_{inv}$ (cm$^{-1}$)\cite{Hutson2009}&$\left|\frac{\mu_{eff}}{\mu}\right|$\\
\hline
1 1&8.26&0.0530&1/2\\
2 2&22.74&0.0525&2/3, 1/3\\
2 1&28.82&0.0520&1/3, 1/6\\
3 3&43.51&0.0518&3/4, 1/2, 1/4\\
3 1&53.60&0.0511&1/4, 1/6, 1/12\\
3 2&59.69&0.0506&1/2, 1/3, 1/6\\
\hline
\end{tabular}
\end{indented}

\end{table}

\section{Experiment}
\begin{figure}
\includegraphics[width = 8.6cm]{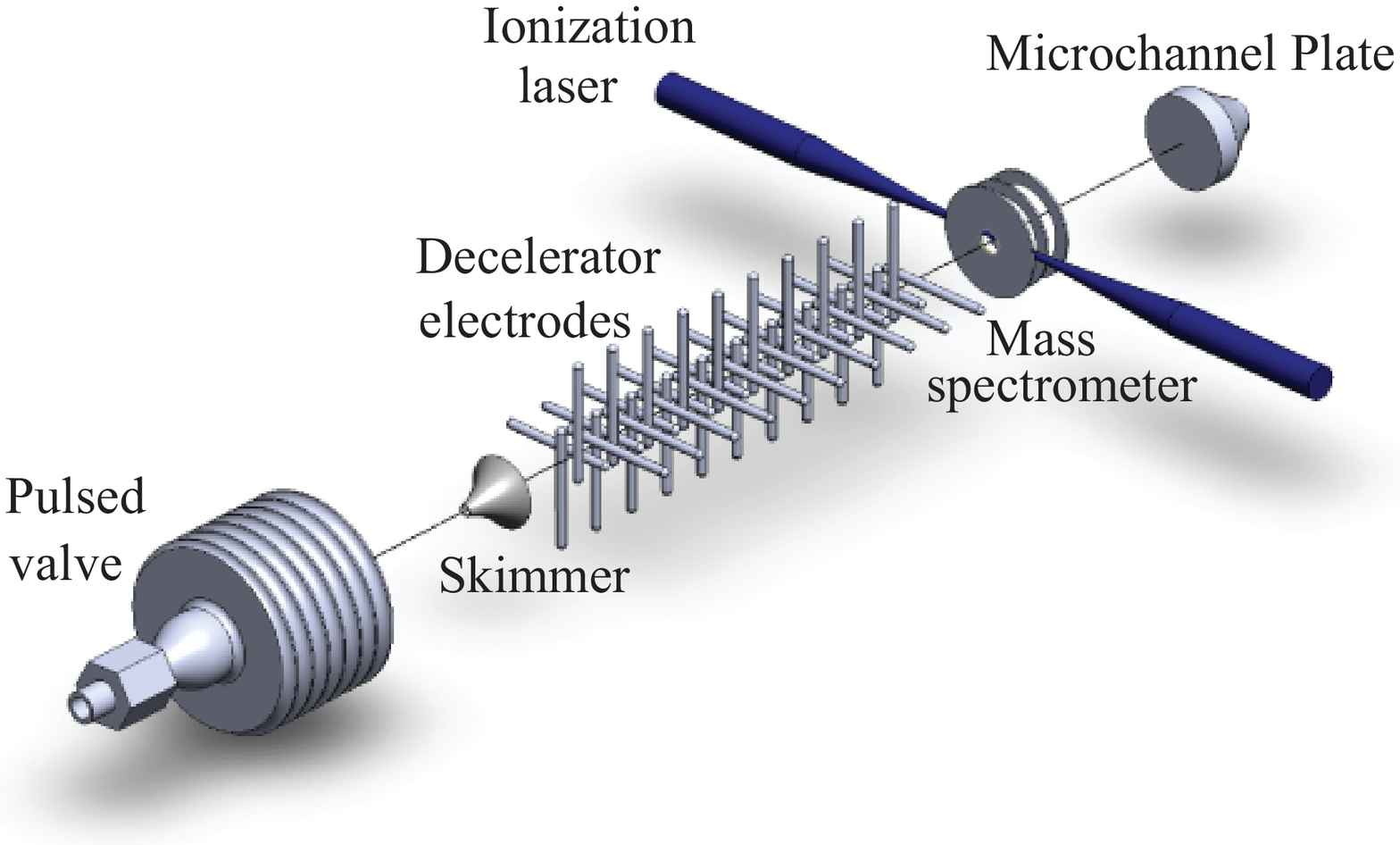}
\caption{Experimental set-up consisting of a PZT-driven pulsed valve, molecular beam skimmer, decelerator, time-of-flight mass spectrometer, and microchannel-plate ion detector.  The decelerator consists of 150 electrode pairs (not all are shown).}
\label{apparatus}
\end{figure}

The experimental apparatus consists of a pulsed valve, Stark decelerator, and molecular detector comprised of a pulsed dye laser, time-of-flight mass spectrometer (TOFMS) and microchannel plate (MCP)(Figure \ref{apparatus}). A molecular beam is created by supersonic expansion of $^{14}$ND$_3$ seeded in krypton via a PZT-actuated pulsed valve. The valve nozzle is constructed from a 2 cm long piece of hypodermic needle tubing and is resistively heated to approximately 185$^\circ$ C to increase the population of molecules in excited rotational states.  The molecular beam, with an initial mean velocity of 460 m/s, passes through a skimmer into a differentially pumped vacuum chamber containing the 149-stage Stark decelerator operating at $\pm 12$kV. After deceleration, the molecules are detected via a $|$J,K$\rangle$ state sensitive $2+1$ resonance enhanced multi-photon ionization (REMPI) scheme near 317 nm.  The ions are then accelerated by the TOFMS onto the MCP.  The signal at the arrival time for an ion with a mass-to-charge ratio of 20 is measured and averaged over many experimental runs.

\begin{figure}
\includegraphics[width = 8.6cm]{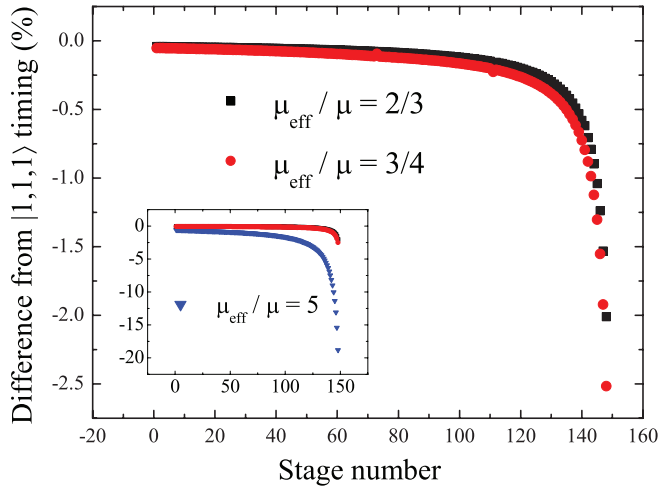}
\caption{Percent differences for the $\frac{\mu_{eff}}{\mu}=\frac{2}{3}$ ($|$2,2,2$\rangle$, squares) and the $\frac{3}{4}$ ($|$3,3,3$\rangle$, circles) timing sequences from the $\frac{\mu_{eff}}{\mu} = \frac{1}{2}$ ($|$1,1,1$\rangle$) for decelerating to 60 m/s.  Inset:  timing difference of the $\frac{\mu_{eff}}{\mu} = 5$.}
\label{timings}
\end{figure}

Operation of the Stark decelerator begins by setting the high-voltage switching timings appropriate for the desired final velocity.  This timing sequence is calculated from trajectory simulations using electric fields generated from Comsol, a finite element analysis package.  To facilitate the discussion of how these timing sequences are related to different rotational states, we introduce the effective dipole moment,
\begin{equation}
\mu_{eff} = \left|\vec{\mu}\right|\frac{M K}{J(J+1)}.
\end{equation}
For the $|$J,K,M$\rangle$=$|$1,1,1$\rangle$, $|$2,2,2$\rangle$, and $|$3,3,3$\rangle$ states, $\frac{\mu_{eff}}{\mu}=\frac{1}{2}$, $\frac{2}{3}$, and $\frac{3}{4}$, respectively. Although $\mu_{eff}$ does not uniquely determine the rotational state of the molecule, it does uniquely determine the timing sequence.  Therefore, to slow synchronous molecules with  different $\mu_{eff}$ values to the same final velocity requires operating the decelerator using different timings.  For example, with our particular apparatus, decelerating a $|$1,1,1$\rangle$ synchronous molecule ($\frac{\mu_{eff}}{\mu}=\frac{1}{2}$, see Table \ref{stark}) from 460 m/s to 60 m/s requires calculating decelerator timings for a phase angle of $\phi_{0}\approx 84^{\circ}$.  Removing the same amount of energy per stage for a synchronous molecule with $\frac{\mu_{eff}}{\mu}=5$ requires a much smaller phase angle of $\phi_{0}\approx 7^{\circ}$.  Calculated switch timings for decelerating the $|$2,2,2$\rangle$ and $|$3,3,3$\rangle$ states to 60 m/s, relative to the $|$1,1,1$\rangle$ state, appear in Figure \ref{timings}.  Even though the effective dipole moments vary from those of the $|$1,1,1$\rangle$ state by as much as 50 \%, the resulting switch timings differ by at most 2.5 \%. The inset in Figure \ref{timings} contains the relative timings for a hypothetical ND$_3$ molecular state with $\frac{\mu_{eff}}{\mu}=5$.

Note that while the Stark shift depends on $M$, our molecular detection scheme is $M$-independent.  Hence, when discussing timings for the decelerator, the full $|$J,K,M$\rangle$ of the synchronous molecule used to calculate the timings is referenced.  Discussion regarding the actual detection of molecules references only $|$J,K$\rangle$, as all possible values of $M$ are included.

When evaluating the state selectivity of the Stark decelerator we consider only the relative populations in weak-field-seeking states.  States that are strong-field-seeking as well as those unperturbed by the electric field will effectively be filtered out.  The former are attracted to the highest electric fields, located at the surface of the electrodes, and therefore do not make it to the end of the decelerator.  The latter are unguided transversely and not decelerated longitudinally and hence make up a portion of the undecelerated background molecular beam.

To test how the final molecule density depends on the chosen timing sequence, we measure relative populations of decelerated rotational states using timing sequences for synchronous molecules with $\frac{\mu_{eff}}{\mu} = \frac{1}{2}, \frac{2}{3},\frac{3}{4}, 2, 3, 4,\textnormal{ and } 5$ decelerated to 60 m/s.  A final velocity of 60 m/s is the lower limit for the $|$1,1,1$\rangle$ synchronous molecule given our beam's initial forward velocity and our number of decelerator stages.  This constitutes a removal of more than 98$\%$ of the initial longitudinal kinetic energy of the molecular beam.  This velocity was chosen for these measurements because a low final velocity results in decelerated molecules that are spatially displaced from the undecelerated background molecular beam.  This displacement effectively amplifies any state selectivity of the decelerator as the undecelerated background molecular beam has a state purity determined solely by the dynamics of the supersonic expansion.

\section{Results and Discussion}

\begin{figure}
\includegraphics[width = 15cm]{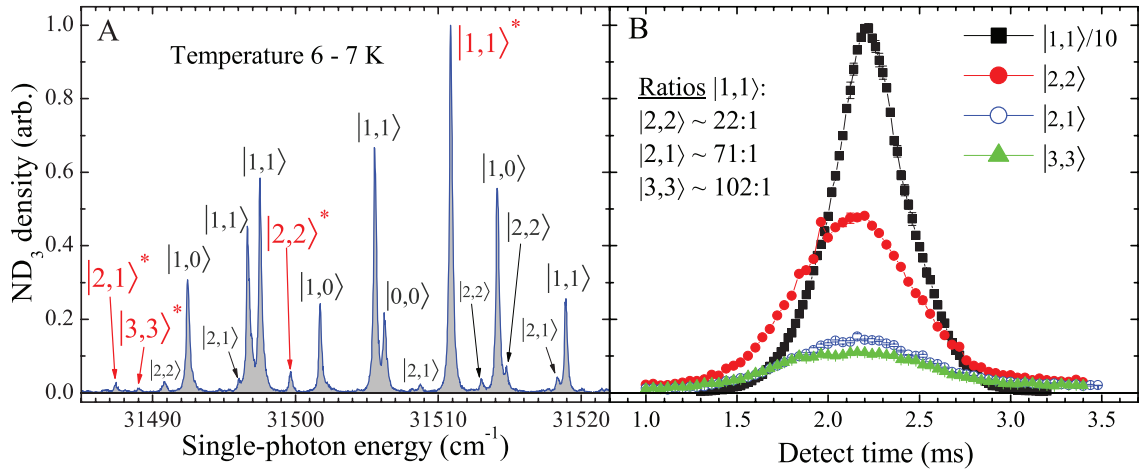}
\caption{Characterizations of the initial molecular beam.  (A) Measured REMPI trace of the molecular beam rotational state distribution where numerous rotational states are populated.  The relative state populations indicate an approximate temperature of 6-7 K.  (B) Time-of-flight signal of various rotational states at the end of the decelerator where no voltage is applied, as detected on the starred wavelengths in (A).  Note that the $|$1,1$\rangle$ state (solid squares) has been divided by 10 for ease of comparison.}
\label{REMPI}
\end{figure}

The distribution of rotational states initially present in the molecular beam was determined using the setup shown in Figure \ref{apparatus}, but with no voltage applied to the decelerator (free flight).  Though not a thermal distribution, the approximate temperature was found to be $6-7$ K by fitting the measured spectrum (Figure \ref{REMPI} (A)) to a theoretical spectrum weighted by Maxwell-Boltzmann factors.  The theoretical spectrum was derived from the Hamiltonian for ND$_3$ using constants together with individual line strength calculations from \cite{Ashfold1987,Bentley2000}.  As shown in Figure \ref{REMPI} (A), the initial molecular beam consists of numerous excited rotational states.  The experimental spectrum has power broadened linewidths of $\sim$ 0.25 cm$^{-1}$, which is narrow enough that, for our chosen detection wavelengths, adjacent lines do not overlap, thus allowing for true single-state detection.  Figure \ref{REMPI} (B) shows the densities of molecules in states $|$1,1$\rangle$, $|$2,2$\rangle$, $|$2,1$\rangle$ and $|$3,3$\rangle$ as a function of detection time under free flight conditions.  The traces have been normalized to the calculated transition line strengths and are therefore on a common vertical scale to facilitate direct comparison. The $|$1,1$\rangle$ trace has also been divided by 10 for display purposes.  Relative rotational-state populations initially present in the molecular beam are quantified using ratios of peak molecule densities with the initial free-flight ratio of $|$1,1$\rangle$:$|$2,2$\rangle$ being $\sim$ 22:1.

\begin{figure}
\includegraphics[width = 16cm]{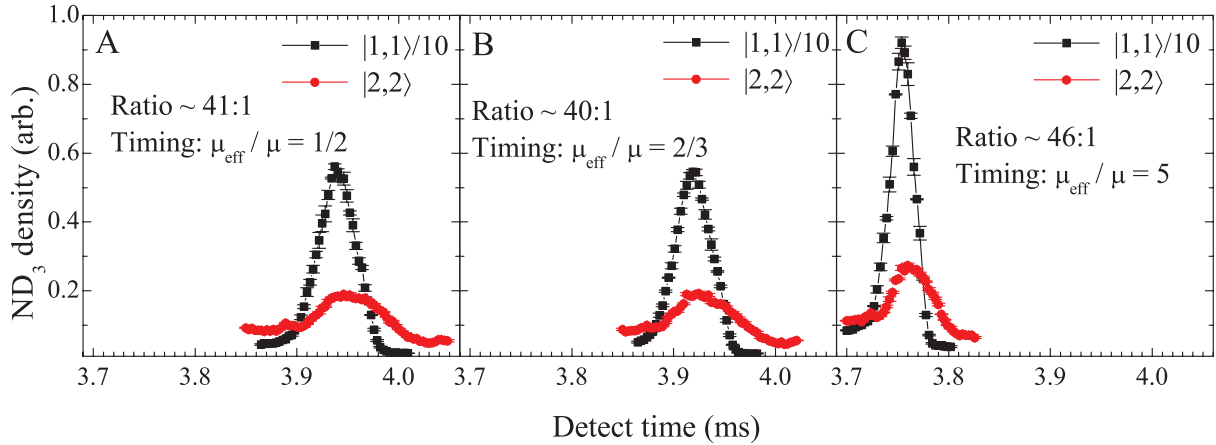}
\caption{
(A) Detecting $|$1,1$\rangle$ (squares) and $|$2,2$\rangle$ (circles) molecules with timings calculated for decelerating a $\frac{\mu_{eff}}{\mu} = \frac{1}{2}$ synchronous molecule to 60 m/s.
(B) Same as A but using a $\frac{\mu_{eff}}{\mu} = \frac{2}{3}$ timing sequence.
(C) Same as A but using an $\frac{\mu_{eff}}{\mu} = 5$ timing sequence.  Note the comparable ratios and the shift in arrival times.}
\label{3panel}
\end{figure}

The ratio of $|$1,1$\rangle$:$|$2,2$\rangle$ is also measured after the deceleration process. If Stark deceleration were capable of generating a molecular beam consisting of a single rotational state, this ratio would exhibit a strong dependence on the timing sequence used.  For instance, the ratio would be maximized when using timings calculated with $\frac{\mu_{eff}}{\mu}=\frac{1}{2}$ (i.e. for a $|$1,1,1$\rangle$ synchronous molecule).  Conversely, this ratio should approach zero when using timings determined with $\frac{\mu_{eff}}{\mu}=\frac{2}{3}$ (i.e. for a $|$2,2,2$\rangle$ synchronous molecule).  Observation of no dependence on the timing sequence would indicate no purification of these two states by the Stark deceleration process.  Measured time-of-flight molecule densities of the $|$1,1$\rangle$ and $|$2,2$\rangle$ states are shown in Figure \ref{3panel} for timings calculated using $\frac{\mu_{eff}}{\mu} = \frac{1}{2}, \frac{2}{3}$, and $5$ synchronous molecules slowed to 60 m/s.  The ratio of the density of $|$1,1$\rangle$ to $|$2,2$\rangle$ molecules does not change appreciably, even when the timing sequence for a dipole moment an order of magnitude larger is used.  Thus, deceleration has little effect on the relative populations of these weak-field-seeking states as no correlation between state purity and timing sequence is observed.

Comparing the ratio of $|$1,1$\rangle$:$|$2,2$\rangle$ between free flight and deceleration using $\frac{\mu_{eff}}{\mu}=\frac{1}{2}$ timing we see an increase of a factor of 2 (from 22:1 to 41:1).  The \emph{same} increase in the $|$1,1$\rangle$:$|$2,2$\rangle$ ratio is seen when using $\frac{\mu_{eff}}{\mu}=\frac{2}{3}$ timing (from 22:1 to 40:1), which indicates the effect is not due to the state selectivity of the deceleration process.  This increase is primarily due to all $M$ sublevels being present in the free-flight beam, but not in the decelerated beam.  Specifically, $1/5$ of the $|$2,2$\rangle$ molecules ($|$2,2,2$\rangle$ only) are decelerated compared to $1/3$ of the $|$1,1$\rangle$ molecules ($|$1,1,1$\rangle$ only).  The weak-field-seeking $|$2,2,1$\rangle$ molecules are absent in the decelerated beam because their Stark shift is too small to be decelerated down to these velocities.  In addition to the $M$ sub-level contribution, the $|$2,2$\rangle$ molecules have a 50 \% wider initial velocity distribution than the $|$1,1$\rangle$ molecules after the supersonic expansion, decreasing the efficiency of the deceleration process and increasing the ratio of $|$1,1$\rangle$:$|$2,2$\rangle$.

\begin{figure}
\includegraphics[width = 13cm]{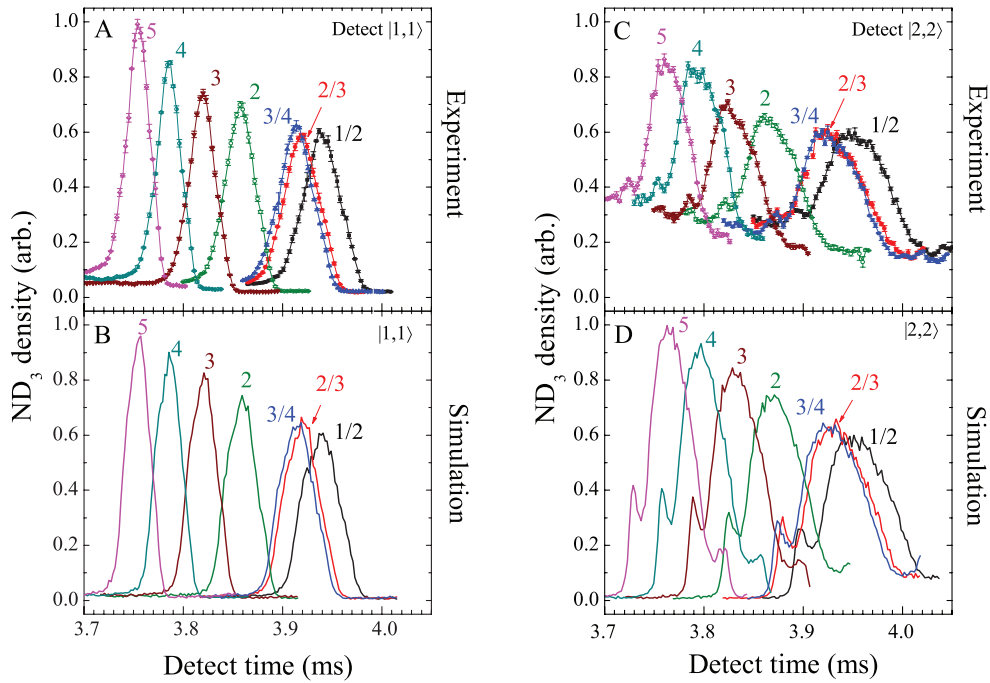}
\caption{ Comparison of experiment and simulation.  Time-of-flight traces are labeled by the corresponding $\frac{\mu_{eff}}{\mu}$ value used to calculate the decelerator timing sequence.  (A) Detecting $|$1,1$\rangle$ using various timing sequences calculated using synchronous molecules with different $\mu_{eff}$ values.  (B) Results from 3D Monte Carlo simulations of the expected $|$1,1$\rangle$ densities. (C) Detecting $|$2,2$\rangle$ under the same conditions as in A. (D) Results from simulations for $|$2,2$\rangle$.}
\label{experimenttheory}
\end{figure}

To further explore the effect of various timing sequences on state selectivity, a single state was detected using timing sequences for $\frac{\mu_{eff}}{\mu} = \frac{1}{2}, \frac{2}{3}, \frac{3}{4}$, 2, 3, 4, and $5$ (Figure \ref{experimenttheory}).  Increasing molecular densities are observed when timing sequences for larger $\frac{\mu_{eff}}{\mu}$ are used, regardless of the state being detected. This behavior can be compared to results from a series of three-dimensional Monte Carlo simulations.  The simulated density as a function of time-of-flight of $|$1,1$\rangle$ and $|$2,2$\rangle$ molecules is shown in Figure \ref{experimenttheory} (B) and (D), respectively.  The simulations correctly predict both the shift in arrival times as well as the increase in molecule density.

\begin{figure}
\includegraphics[width = 15cm]{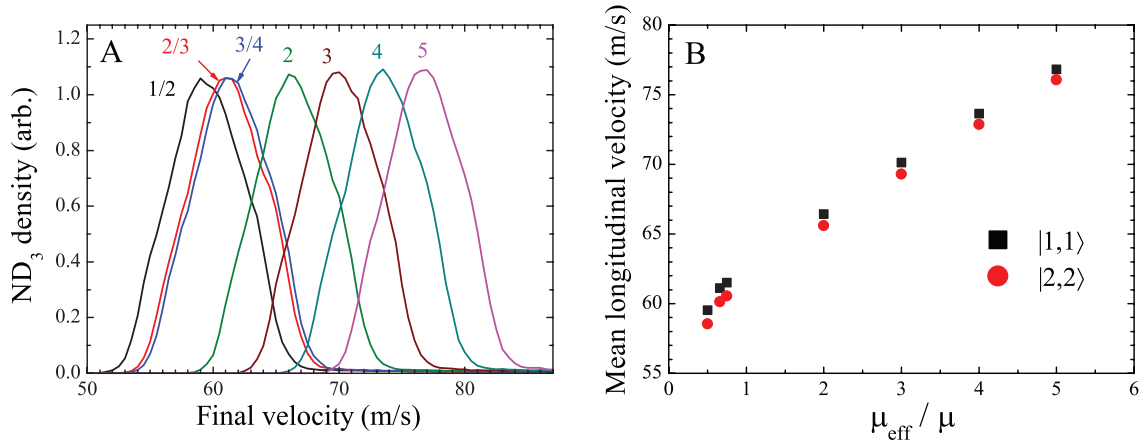}
\caption{Resulting final velocity distribution for molecules in the $|$1,1$\rangle$ state using various $\frac{\mu_{eff}}{\mu}$-value timing sequences as produced by Monte Carlo simulations.  (A) Curves are labeled by the corresponding $\frac{\mu_{eff}}{\mu}$ value used to calculate the decelerator timing sequence. (B) Mean longitudinal velocities extracted from simulations.}
\label{velocity}
\end{figure}

The simulations were also used to extract information about the final beam that is difficult to determine experimentally.  Of particular interest are the final molecular velocity distributions at the exit of the decelerator under different timing schemes (Figure \ref{velocity}).  As shown, decelerator timing sequences based on increasing $\frac{\mu_{eff}}{\mu}$ result in increasing final velocities.  This shift in velocity, together with the timing difference of the last stage (Figure \ref{timings}), combine to produce the observed shift in the arrival times (Figure \ref{experimenttheory}).  Figure \ref{velocity} (B) shows the final mean longitudinal velocity for both $|$1,1$\rangle$ molecules and $|$2,2$\rangle$ molecules.  The resulting final velocities for these two states are nearly identical under all timing conditions.  In a single realization of the experiment, only one timing sequence can be used, resulting in a molecular beam consisting of both $|$1,1$\rangle$ and $|$2,2$\rangle$ molecules with nearly identical final velocities and spatial distributions.  Thus both weak-field-seeking states will be present in the resulting decelerated molecular beam.

Studying gas-phase molecular interactions where the reactants are in a single internal quantum state and have well-controlled external degrees of freedom is a major goal in physical chemistry. Stark deceleration is a possible candidate for producing these single-state molecular beams as it is a proven method for controlling external degrees of freedom. In addition, Stark deceleration is capable of filtering out strong-field-seeking states, states that are unperturbed by the electric field, and weak-field-seeking states with Stark shifts too weak to be slowed to the desired final velocity, making it an increasingly viable candidate.  Our results show that in many circumstances Stark deceleration is not capable of filtering out all but one of the remaining weak-field-seeking states.  In fact, every weak-field-seeking state present in the supersonic expansion that can be decelerated to the desired final velocity, will be decelerated regardless of whether its Stark shift is stronger or weaker than that of the synchronous molecule.  This is an important consideration when interpreting results of collision experiments using Stark deceleration and analogous techniques.

\ack
This work was supported by NSF (PHY 0551010 and PHY 0748742), Air Force Office of Scientific Research (FA9550-08-1-0193 and FA9550-09-1-0588), and the Alfred P. Sloan Foundation.  The authors would like to thank Jun Ye and Ben Stuhl for helpful discussions.

\end{document}